# Causes of Excess Capacity

Samidh Pal[1] [2] [3]

ORCID ID: https://orcid.org/0000-0001-9227-6431

*Abstract:*
*This study delves into the origins of excess capacity by examining the reactions of capital, labor, and capital intensity. To achieve this, we have employed a novel three-layered production function model, estimating the elasticity of substitution between capital and labor as a nested layer, alongside capital intensity, for all industry groups. We have then selectively analyzed a few industry groups for comparative purposes, taking into account the current government policies and manufacturing plant realities. Ultimately, we recommend that policymakers address the issue of excess capacity by stimulating the expansion of manufacturing plants with cutting-edge machinery. Our findings and recommendations are intended to appeal to academics and policymakers alike.*

*JEL Code:* *L6, L5.*

*Keywords:* *Optimization Algorithms, Constant Elasticity of Substitution, Capital Intensity, Technical Efficiency, Capital, Labor.*

[1] *Pursuing Ph.D, Department of Economic Sciences, University of Warsaw, Poland.* *s.pal@uw.edu.pl*;

[2] **Statements and Declarations:** The authors of this research paper declare that they have no financial or personal relationships that could inappropriately influence or bias their work. They also have no professional or personal biases that could affect the research findings or interpretation of the results. The authors have no conflicts of interest to report.

[3] **Competing Interests:** The authors declare that they have no known competing financial interests or personal relationships that could have appeared to influence the work reported in this paper.

The concept of excess capacity in industrial enterprises is a well-known phenomenon that is often difficult to achieve in practice. Manufacturing industries typically operate at less than ideal capacity, resulting in excess capacity that is distinct from actual capacity (Perelman, 1989, p. 298). This excess capacity can impede the development of new branches within a firm and lead to the formation of oligopolistic or monopolistic market structures.

In standard neoclassical analysis, excess capacity is often present in monopolistically competitive firms (Todorova, 2015), while in imperfectly competitive markets, manufacturing plants must operate at a certain level of production capacity. Excess capacity can occur when market demand falls below the level required to sustain the installed production capacity (Jensen, 1999).

Excess capacity acts as a barrier to entry in an industry and impedes the growth of new units (Wenders, 1971) (Pashigian, 1968). Despite flexible prices and positive interest rates, some suppliers along the production chain operate under fixed costs, and the theory of economic slack is based on firms that face only fixed costs over a range of outputs. This theory explains the procyclicality of firm entry and capacity utilization (Murphy, 2017).

Our study aims to identify the problem of excess capacity and its effects on industrial performance. We expect to observe greater excess capacity in an industry group when the elasticity of substitution between capital and labor, as well as capital intensity, is one. This indicates an opportunity to expand capital intensity as substitutes to reduce the cost of labor and capital.

Pal (2019) has shown that for industrialization, a region requires capital investment, highly skilled labor, and rapid development of infrastructure. He also studied the inter-regional intra-industry disparity in India's West Bengal region and found that it is facing a crisis due to underutilization of capacity and poor industrial performance. This is attributed to the lower level of capital intensity and average efficiency of labor and capital, leading to a waste of capital and a persistent problem of poor industrial performance in the region.

In conclusion, excess capacity in industrial enterprises is a complex issue that can have significant effects on industrial performance. Our study aims to identify the factors contributing to excess capacity and its effects on industrial performance. Further research is needed to fully understand the effects of excess capacity on industrial performance and to identify potential solutions to this problem.

**Literature review**
In recent years, several Indian economists have attempted to identify the problem of excess capacity in Indian industries. However, their methodologies and findings have varied. In this literature review, we have collected and analyzed some of their studies to shed light on this issue.

Budin and Paul (1961) used time series data to measure capacity utilization in the Consumer Group, Intermediate Group, and Infrastructural Group. Their findings indicated that excess capacity was present, but they interpreted it as being very low. Alagh and Shah (1972) collected input and output data to compare potential, actual, and excess capacity in 16 major industries. They found that excess capacity was more significant than potential production capacity in the transport equipment and nonelectrical equipment sector, while between one-third and half of potential production capacity was utilized in other industries.

Srinivasan (1992) conducted a basic level of research on capacity utilization in the same three primary industries. Their study found that 25% to 30% excess capacity prevailed in these three sectors during the period of 1970 to 1984. Kumar and Arora (2009) used time series data to analyze capacity utilization in the Indian Sugar Industry from 1974-1975 to 2004-2005. They found that there was approximately 13% excess capacity in the industry and suggested that more labor and intermediate inputs were required to achieve full capacity.

Ray (2011) analyzed the level of capacity utilization efficiency in the Indian rubber industry from 1979-1980 to 2008-2009. Their study revealed a declining growth rate of capacity utilization in the industry during the post-reform period, accompanied by declining output growth as well as capacity growth. They suggested that the low correlation between capacity expansion and lagged output indicated unintended excess capacity in the Rubber Industry, which varied from year to year.

While these studies support the existence of excess capacity in Indian industries, none of them measured the actual causes of excess capacity except for Arora. Furthermore, none of them discussed or mentioned the effects of factors used in production that can be responsible for excess capacity.

To address this gap, our study used a comparative research methodology and performed a nonlinear optimization algorithm for our new nested production function. Our findings shed new light on the causes of excess capacity in Indian industries and offer insights for policymakers and practitioners.

**Methodology review**
Our new nested production function by Pal (2023) is exactly formed as Sato's (1967) nested CES function.

$$V = A\left[\delta(\delta_1 K^{-\rho_1} + (1-\delta_1)L^{-\rho_1})^{\frac{\rho}{\rho_1}} + (1-\delta)\left(\frac{K}{L}\right)^{-\rho}\right]^{-\frac{1}{\rho}} \qquad (1)$$

The validation of the new production function model (eq. 1) involves analyzing the concavity of the function. We study the marginal products of labor and capital in our nested production function model, and if $\rho_1 = \rho$, the function reverts to the plain n-input CES function.

Our production function considers three inputs: capital (K), labor (L), and a composite input (K/L). If the elasticities of substitution between K and L and the elasticity of substitution between (K, L) and composite input K/L are identical, then the nested technologies are returns to Cobb-Douglas form. This means that the model is more generalized, as we can focus on the elasticity of substitution between (K, L) and composite input K/L, which will give the theoretically reasonable parameter (i.e. $\sigma \approx [0,1]$) to understand the problem of excess capacity.

The parameters in the model, including $A, \delta, \delta_1, \rho$, and $\rho_1$ have specific meanings.[4] For instance, if $\delta = 0$, it means that the production function only depends on the capital and labor inputs. If $\delta = 1$, it means that the production function only depends on the composite input. If $\delta_1 = 0$, it means that the composite input is purely a function of labor. If $\delta_1 = 1$, it means that the composite input is purely a function of capital.

In order to determine whether an industry is capital or labor intensive, we can utilize the $\delta$, and $\delta_1$ parameters. A $\delta$ value of 1 and a $\delta_1$ value of 1 indicate a purely capital-intensive industry, while any other combination suggests a labor intensive industry. By utilizing these parameters, we can accurately measure the level of capital or labor required in a particular industry. This approach is particularly useful for academics and researchers seeking to understand the economic dynamics of various industries.

In the 1960s and 1970s, the nonlinear least-square estimation methods in statistical analysis faced challenges. Transforming the nonlinear three-input nested CES function into a linearized form was difficult. Directly estimating the nested function through the use of the 'nls' function in R, which is a nonlinear least squares estimation method, has proven to be insufficient in many real-world data applications. This is because convergence is often not achieved or the estimated parameters are theoretically unreasonable.

As an alternative, we consider using the Levenberg-Marquardt nonlinear optimization algorithm (1963) for our nonlinear three-input nested production function. This algorithm's aim is to either achieve convergence or estimate theoretically reasonable parameters. The main objective of this optimization algorithm is to minimize the sum of squared residual errors for parameters.

To ensure convergence, we have set a maximum number of iterations for this algorithm. The purpose of this study is to identify the causes of excess capacity.

The Levenberg-Marquardt algorithm (1963) is the most popular and most commonly used to estimate the parameters of the CES function. This algorithm performs with an optimum level of interpolation between the first-order Taylor series expansion and the gradient method algorithm. Furthermore, it can be seen as a maximum neighborhood method. Within these two kinds of non-linear estimation techniques, it reduces the weaknesses and increases the convergence limit. However, this algorithm still performs little weakly, as it also estimates the elasticity of substitution that becomes biased towards infinity, zero, or unity.

Optimizing complex functions is a challenge, particularly when the minimum region presents a flat surface of negative squared residuals. This is especially true for n-input nested functions. However, the grid search method offers a reliable and efficient approach to estimate the CES function. By pre-determining the value of $\rho$, we can increase the accuracy of the remaining parameters, limiting them to a specific surface.

---

[4] The parameter $A$ represents the level of technology and technological efficiency, which is assumed to be positive. The parameter $\delta$ determines the weight given to the composite input (K/L) in the production function. The parameter $\delta_1$ determines the weight given to the capital input (K) in the composite input (K/L). The parameter $\rho$ determines the curvature of the production function and the parameter $\rho_1$ determines the curvature of the composite input.

Despite its advantages, conducting a grid search examination of our new nested function presents difficulties. Selecting an appropriate grid value for $\rho$ is a significant challenge, and existing software and estimation techniques do not aid in determining the optimal grid value. Therefore, the best grid value must be selected through graphical analysis, requiring a careful and thorough approach.

**Data**

We utilized the Annual Survey of Industry (ASI) data (Central Statistical Organization, 2013/2019) from the Central Statistical Organization, (2008) (CSO) for seven years, covering all registered manufacturing industries in six major industrial states. The data was based on the National Industrial Classification (NIC) system, which consists of three-digit codes. Our analysis focused on three inputs: Capital, Labor, and Capital intensity.

**Selection of model**

The elasticity of substitution between Capital and Labor is a crucial parameter in economic analysis, with a range of $\sigma \approx [0,1]$ considered theoretically reasonable (Bui Khac, et al., 2018). To estimate our nested production function, we employed a grid search with the Levenberg-Marquardt nonlinear optimization algorithm, fixing the value of $\rho$ while exploring two different grids of $\rho$ values. The preselected $\rho$ grids are: *rhoVec_1 = [(-0.9, 1.25, 0.64), (1.68, 1.72, 0.88), (1.86, 10.00, 0.94)] and for rhoVec_2 = [(-1, 1, 0.40), (1.68, 2.00, 0.64), (10.00, 11.51, 0.88)]*. The use of nonlinear least-squares estimation methods can result in "flat surfaces" around the minimum, particularly with broad ranges of substitution parameters ($\rho_1$ and $\rho$). This challenge can be overcome through the use of a grid search, which allows for exploration of the substitution parameters and identification of the minimum objective function. By employing this approach, we were able to accurately estimate our nested production function for each industry group.

| Figure: 1 | Figure: 2 |
|---|---|
| Negative sum squared residuals. | Negative sum squared residuals. |
| Industry Code: 151 (with *rhoVec_1*) | Industry Code: 251 (with *rhoVec_2*) |
| 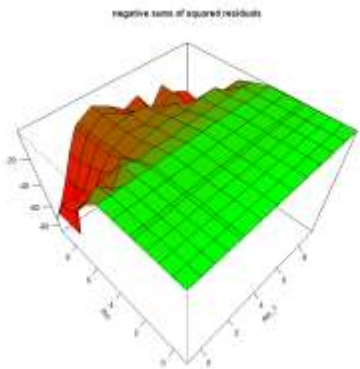 | 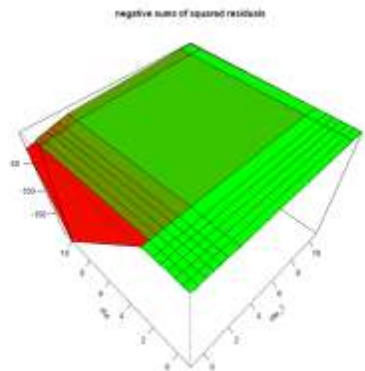 |

Source: Calculated by the author from Annual Survey of Industries (ASI) data 2016-17.

The evaluation of replacement parameters ($\rho_1$ and $\rho$) is carried out using a pre-selected grid of values, while the remaining parameters are estimated through nonlinear least-squares estimation. This process can be performed in one, two, or three dimensions. In figure: 1 and 2 are utilized to facilitate the identification of the highest and lowest parts of the area by representing the negative sum of squared residuals against $\rho_1$ and $\rho$.

To compare two industry groups, 151 (*Manufacture of Leather Products*) and 251 (*Manufacture of Structural Heavy Metal Products*), a random selection was made. The analysis revealed that for industry group 151, the small sum of squared residuals $\rho$ is represented by the green part of the plot, while the large sum of squared residuals $\rho_1$ is represented by the red part of the plot when using the *rhoVec_1* model in figure: 1. The model fits best when the interval of $\rho$ is [-1, 1] and the interval of $\rho_1$ is approximately [0.94, 9], with $\rho$ being smaller than 9. However, if $\rho$ is larger than 8 and $\rho_1$ has a low value (with an upper limit between -0.90 and 0.94 depending on the value of $\rho$), the model fit is at its worst. A similar trend was observed for industry group 251 and the *rhoVec_2* model in figure: 2. The interval of $\rho_1$ is approximately [0.88, 10] and the model fits best when $\rho$ is smaller than 10. However, if $\rho$ is larger than 6 and $\rho_1$ has a low value (with an upper limit between -1 and 0.88 depending on the value of $\rho$), the model fit is at its worst.

Based on these results, it can be concluded that $\rho_1$ is not equal to $\rho$, and as a result, our nested model is not a return to the n-input CES model. Therefore, in our nested model, the third nested input (the capital-labor ratio) acts as a technology in and of itself. Furthermore, the elasticity of substitution values for these two industry groups and their respective rho sets were estimated to be theoretically reasonable parameters.

**Empirical result and interpretation**

Table: 1

CES estimation with grid search model for 151 and 251 industries

| Rho_set | $R^2 = 1$ | StdError | Elasticity of Substitution (AU) ($\sigma = [0,1]$) | $\delta$ | $\delta_1$ | RSS | Convergence |
|---|---|---|---|---|---|---|---|
| Industry Code: 151 | | | | | | | |
| rhoVec_1 | 0.99 | 0.43 | 1.4 ≈ 1 | 1.13 | 0.94 | 0.30 | Achieved |
| Industry Code: 251 | | | | | | | |
| rhoVec_2 | 0.96 | 0.26 | 0.5 ≈ 0 | 1.01 | 1.05 | 1.42 | Achieved |

Source: Calculated by the author from Annual Survey of Industries (ASI) data 2016-17.

Upon analysis of table: 1 the values of $\delta$ and $\delta_1$, it is evident that the industry groups 151 and 251 are purely capital intensive. The estimated elasticity of substitution for these industry groups is approximately 1.4 and 0.5 respectively, indicating that our production function is appropriate for both groups. These values fall within the range between one and zero, highlighting the appropriateness of theoretically reasonable.

Industry group 151 exhibits excess capacity, indicating a lack of incentive to adopt advanced machinery to address the shortage of high-skilled labor and capital. On the other hand, the substitution elasticity for industry group 251 suggests a weak substitutability between the factors of production, implying that there is no need to expand the production unit with new machinery to overcome the shortage of labor and capital.

**Conclusion**

The leather product manufacturing industry group 151 is dominated by a few units that produce similar products with limited local demand. This is due to consumers' preference for higher quality leather products from outside the region, as local products are considered inferior in terms of quality and fashion. Consequently, these units experience a decline in market share. In contrast, units in other regions that offer both quality and variety of leather products have greater market access. For instance, in West Bengal, there is a range of medium, small, and micro units that produce diverse leather products and enjoy a considerable market share outside the state. However, the overall efficiency of the leather product manufacturing industry group in West Bengal remains low.

The comparative study in figure 3 reveals that industry group 151 has a lower overall output compared to industry group 251, with all states contributing to the higher output of industry group 251. To improve the performance of the leather product manufacturing industry group 151, there is a need to focus on quality and variety to meet consumers' demands. Moreover, there is a need to enhance the efficiency of the production process to increase output and competitiveness. By doing so, the industry group can expand its market share and contribute to the economic growth of the region.

Figure: 3
Comparative state-wise total output analysis between two industry groups.
(2016-17) (Rs. In millions)

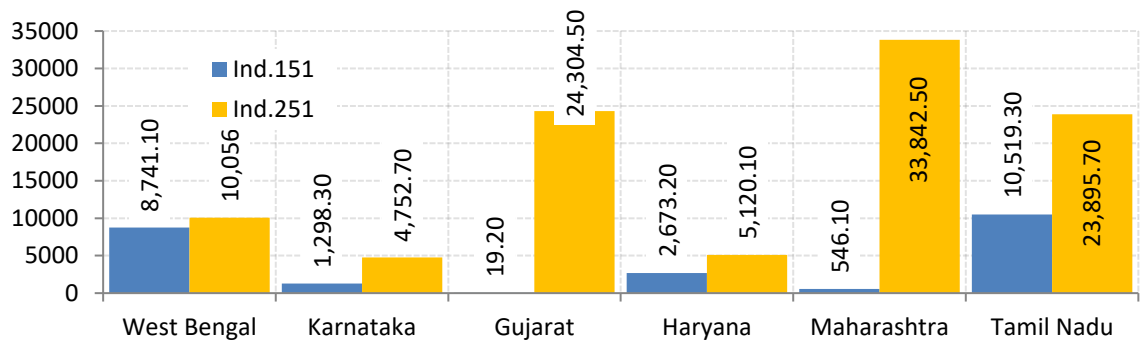

Source: Calculated by the author from Annual Survey of Industries (ASI) data 2016-17.

The manufacturing facilities in West Bengal's industry group 151 have not undergone expansion in the state over the past decade, despite the use of machinery for production. However, our findings suggest that this industry group has significant potential to increase its capital intensity to compensate for the shortages of labor and capital.

Our spatial analysis (Map: 1) reveals that all the manufacturing facilities for industry group 151 are located in the southern region and are concentrated in a single area. This indicates that the state government's policies are not evenly distributed across all sub-regions of the state, or that the existing manufacturers are not aware of these policies and schemes. Consequently, investors are still not ready to invest in the region, and this industry group remains underdeveloped, limiting employment opportunities in the state.

To address this issue, it is crucial to distribute policies and schemes evenly across all sub-regions of the state and raise awareness among existing manufacturers. This will encourage investors to invest in the region and promote the development of industry group 151. By doing so, the state can create more employment opportunities and contribute to the growth of the manufacturing sector.

Map: 1

Plots of 151 industrial units of West Bengal coordinate wise through spatial data analysis.

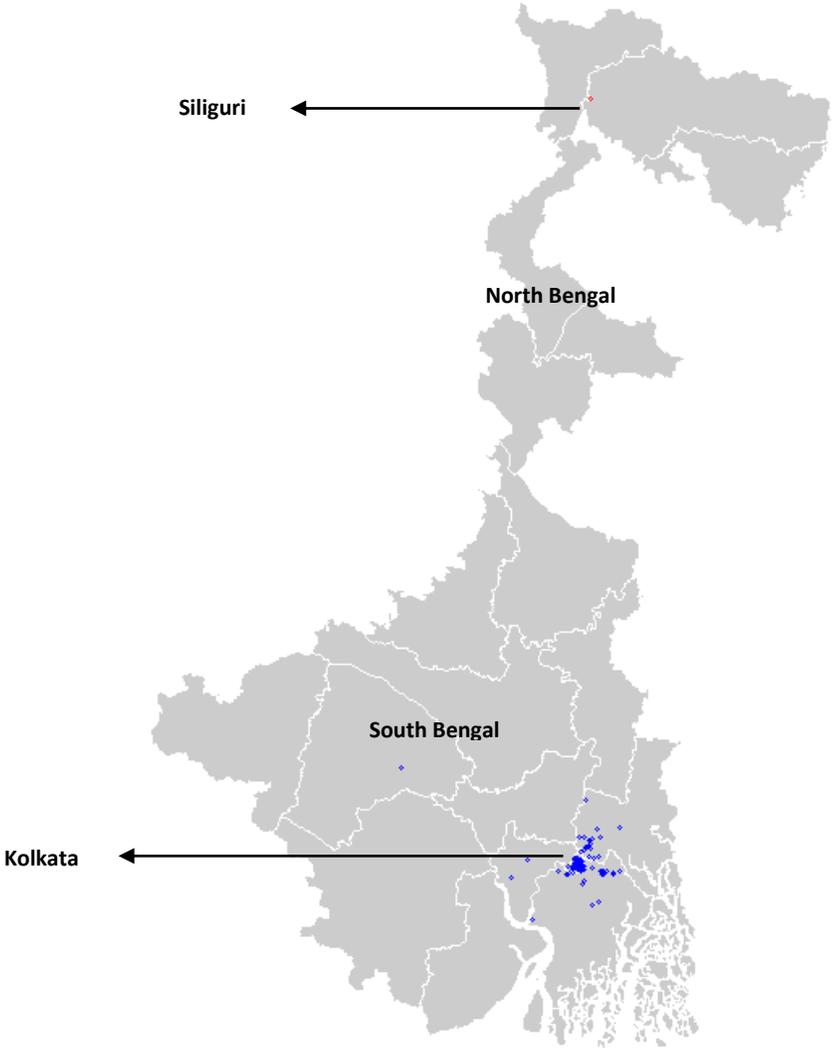

Source: plotted by author from (ASI, 2016-17).

The Federation of Indian Chambers of Commerce and Industry (FICCI) (2010) recently published a report on the Indian manufacturing industry, which revealed that the sector's capacity utilization remains suboptimal. This is likely due to a lack of access to updated technology or insufficient opportunities to upgrade existing technology. The technology gap between sub-industries continues to widen, exacerbating the issue of excess capacity. Moreover, manufacturing plants in West Bengal often face shortages of working capital, leading to a decline in their operational activity.

In a free-market economy, the ability to minimize production costs through advanced machinery provides an advantage to manufacturers operating in other regions. These manufacturers supply their products through intermediaries in the markets of underdeveloped regions, putting pressure on local medium, small, and micro units. This competition fear hinders the growth and expansion of existing and new manufacturing plants in these regions.

To address these challenges, manufacturing units in West Bengal require efficient technology to produce high-quality and diverse advanced and updated products. By doing so, they can remain competitive in the global market and attract more investment. It is essential to bridge the technology gap and provide adequate working capital to enable the manufacturing industry to thrive and contribute to India's economic growth.

**Policies and Suggestion**

Various initiatives and subsidies have been implemented by the central and state governments to support industrial units in upgrading and expanding their manufacturing facilities. The Credit Linked Capital Subsidy Scheme (CLCSS) has been implemented by the Development Commissioner, Ministry of Micro, Small, and Medium Enterprises (MSME), to encourage production technology upgrades in these units, providing an upfront capital subsidy to aid the adoption of advanced technology.

In West Bengal, the Department of Information Technology and Electronics (2018), Information Technology and Electronics Policy has been introduced to establish the state as a leading producer of globally competitive advanced machinery, with the vision of transforming the state into a technologically advanced society and creating more employment opportunities for the youth, as well as becoming a preferred investment destination for investors.

However, despite these efforts, the industry remains concentrated only in the southern region, with low investment in other parts of the state. To address this, regular surveys should be conducted by policy makers to assess the level of productive efficiency in industries and distribute aid promptly to address any excess capacity. Additionally, the capital intensity requirements of each industry should be analyzed, and investment should be prioritized to increase capital-intensive production, thereby reducing potential excess capacity and encouraging growth in the manufacturing sector.

**References:**
Alagh, Y. and Shah, J. (1972) Utilization of Industrial Capacity with Agricultural Growth: A Numerical Exercise for Short-Run Policy. Economic and Political Weekly, 7(5/7), 379-384.
Budin, M. and Paul, S. (1961) The Utilization of Indian Industrial Capacity (1949-1959). Indian Economic Journal, 19(1). https://www.proquest.com/openview/631a7cf111b10e3a8b9349b76969e72d/1?pq-origsite=gscholarandcbl=1819379
Bui Khac, L., Hoang Thi Nhat, H. and Bui Thanh, H. (2018) Factor substitution in rice production function: the case of Vietnam. Economic Research-Ekonomska Istraživanja, 31(1), 1807–1825. https://doi.org/10.1080/1331677x.2018.1515643
Department of Information Technology and Electronics (2018) West Bengal Information Technology and Electronic Policy. Government of West Bengal, Kolkata. https://static.investindia.gov.in/s3fs-public/2019-04/wb-itpolicy-book2018.pdf
Federation of Indian Chambers of Commerce and Industry (2010) FICCI Quarterly Survey on Indian Manufacturing Sector. FCCI [Report]. http://ficci.in/Sedocument/20075/FICCI-Manufacturing-Survey-July-Sept-2010.pdf
Jensen, M. C. (1999) Modern Industrial Revolution, Exit, and the Failure of Internal Control Systems. SSRN Electronic Journal. https://doi.org/10.2139/ssrn.93988
Kumar, S. and Arora, N. (2009). Analyzing Regional Variations in Capacity Utilization of Indian Sugar Industry using Non-parametric Frontier Technique. Eurasian Journal of Business and Economics, 2(4), 1-26. https://www.ejbe.org/index.php/EJBE/article/view/21
Marquardt, D. (1963) An Algorithm for Least-Squares Estimation of Nonlinear Parameters. Journal of the Society for Industrial and Applied Mathematics, 11(2), 431-441.
Murphy, D. (2017) Excess capacity in a fixed-cost economy. European Economic Review, 91, 245–260. https://doi.org/10.1016/j.euroecorev.2016.11.002
Pal, S. (2019) Measuring the Industrial Concentration and Regional Specialization of Major Indian Industrial States. The Indian Economic Journal, 67(3–4), 216–232. https://doi.org/10.1177/0019466220946330
Pal, S. (2023) A New Production Function Approach. arXiv, arXiv:2303.14428 [econ.TH]. https://doi.org/10.48550/arXiv.2303.14428
Pashigian, B. P. (1968) Limit Price and the Market Share of the Leading Firm. The Journal of Industrial Economics, 16(3), 165. https://doi.org/10.2307/2097557
Perelman, M. (1989) Keynes, Investment Theory and the Economic Slowdown: The Role of Replacement Investment and q-Ratios (1989th ed.). Palgrave Macmillan.
Ray, S. P. (2011) Measuring Capacity Utilization and Evaluating the Impact of Liberalization on Capacity Utilization of Indian Drug and Pharmaceutical Industry. Journal of Emerging Knowledge on Emerging Markets, 3(1). https://doi.org/10.7885/1946-651x.1045
Sato, K. (1967) A Two-Level Constant-Elasticity-of-Substitution Production Function. The Review of Economic Studies, 43(2), 201–218. http://www.jstor.org/stable/2296809.
Srinivasan, P. V. (1992) Excess Capacities in Indian Industries: Supply or Demand Determined? Economic and Political Weekly, 27(45), 2437-2441.
Todorova, T. (2015) Is There Excess Capacity Really? Theoretical and Practical Research in Economic Fields, VI, 127-143, 10.14505/tpref. https://doi.org/10.14505/tpref
Wenders, J. T. (1971) Excess Capacity as a Barrier to Entry. The Journal of Industrial Economics, 20(1), 14. https://doi.org/10.2307/2098281